\documentclass[12pt]{article}

\usepackage{amsmath}
\usepackage{amssymb}
\usepackage{amstext}
\usepackage{amscd}
\usepackage{amsfonts}
\usepackage{graphicx}
\usepackage{color}

\begin{document}
\newcommand{\nd}{\noindent}
\newcommand{\nl}{\newline}
\newcommand{\be}{\begin{equation}}
\newcommand{\ee}{\end{equation}}
\newcommand{\ben}{\begin{eqnarray}}
\newcommand{\een}{\end{eqnarray}}
\newcommand{\nn}{\nonumber \\}
\newcommand{\ii}{\'{\i}}
\newcommand{\pp}{\prime}
\newcommand{\expq}{e_q}
\newcommand{\lnq}{\ln_q}
\newcommand{\quno}{q-1}
\newcommand{\qunoinv}{\frac{1}{q-1}}
\newcommand{\tr}{{\mathrm{Tr}}}
\newcommand{\1}{\beta}
\newcommand{\2}{\gamma}
\newcommand{\3}{\rho}

\title{\small{{\bf STATISTICAL MECHANICS OF PHASE-SPACE CURVES}}}

\author{{\small M. Rocca$^1$,   A. Plastino$^1$, and G. L. Ferri$^2$} \\
\small{$^1$ La Plata National University and
  Argentina's National Research Council}\\
\small{(IFLP-CCT-CONICET)-C. C. 727, 1900 La Plata - Argentina}\\
$^2$ \small{Fac. de C. Exactas-National University La Pampa,} \\
\small{Peru y Uruguay, Santa Rosa, La Pampa, Argentina}}

\date{March 20, 2013}

\def\baselinestretch{2}

\maketitle

\begin{abstract}
\nd
     We study the classical statistical mechanics of a phase-space curve.
     This unveils a mechanism that, via the associated entropic  force, provides us with a
      simple realization of effects such as  confinement, hard core, and asymptotic
      freedom.
      Additionally, we obtain negative specific heats, a
      distinctive feature of self-gravitating systems and negative pressures, typical of dark energy.

\nd {\small KEYWORDS: Phase-space curves, Entropic force,
Confinement, Hard core, Asymptotic freedom, Self-gravitating
systems}.

\end{abstract}

\newpage

\renewcommand{\theequation}{\arabic{section}.\arabic{equation}}

\section{Introduction}

\nd We will study here the classical statistical mechanics of
arbitrary phase-space curves $\Gamma$ and unveil some interesting
effects, like confinement and hard-cores. Remind that by
confinement one understands the physics phenomenon that impedes
isolation of color charged particles (such as quarks), that cannot
be isolated singularly. Therefore, they cannot be directly
observed. In turn, asymptotic freedom is a property of some gauge
theories that causes bonds between particles to become
asymptotically weaker as distance decreases. Finally, in the case
of a ``hard core'' repulsive model, each particle (usually
molecules, atoms, or nucleons) consists of a hard core with an
infinite repulsive potential. \vskip 3mm

\nd Our curves-analysis will  provide, in classical fashion, a
simple entropic {\bf mechanism} for these three phenomena.  The
so-called entropic force is a {\it phenomenological} one arising
from some systems' statistical tendency to increase their entropy
\cite{pol1,pol2,verlinde,path,dewar}. No appeal is made to any
particular underlying microscopic interaction.  The text-book
example is the elasticity of a freely-jointed polymer molecule
  (see, for instance, \cite{pol1,pol2} and references therein).
However, Verlinde has argued that gravity can also be understood
as an entropic force \cite{verlinde}. Same for the Coulomb force
\cite{wang}, etc. For instance, we have an exact solution for the
static force between two black holes at the turning points in
their binary motion \cite{uno} or investigations concerning the
entanglement entropy of two black holes and an associated
entanglement entropic force \cite{dos}. A causal path entropy
(causal entropic forces) has been recently appealed to for links
between intelligence and entropy \cite{path}. \vskip 2mm

\nd Here we appeal to an extremely simple model to show that
confinement can be shown to arise from entropic forces. Our model
involves a quadratic Hamiltonian in phase-space. \vskip 2mm

\nd Quadratic Hamiltonians are well known both in classical
mechanics and in quantum mechanics. In particular, for them the
correspondence between classical and quantum mechanics is exact.
However,  explicit formulas are not always  trivial. Moreover, a
good knowledge of quadratic Hamiltonians is useful in the study of
more general quantum Hamiltonians (and their associated
Schroedinger equations) for the semiclassical regime. Quadratic
Hamiltonians are also important in partial differential equations,
because they give non trivial examples of wave propagation
phenomena. Quadratic Hamiltonians are also of utility because they
help to understand properties of more complicated Hamiltonians
used in quantum theory.

\nd We wish here to appeal to quadratic Hamiltonians in a
classical context in order to discern interesting whether some
interesting features are revealed concerning the entropic force
along phase-space curves. We will see that the answer is in the
affirmative.

\section{Preliminaries}

We consider a typical,  harmonic oscillator-like Hamiltonian in
thermal contact with a heat-bath at the inverse temperature
$\beta$.

 \be
\label{ep1.1}    H(p,q)= p^2+ q^2, \ee where $p$ and $q$ have the
same dimensions (that of $H$, obviously).  The corresponding
partition function is given by \cite{patria,lavenda,katz}
\[Z(\beta)=\int\limits_{-\infty}^{\infty}\,e^{-\beta
H(p,q)}\;dpdq=\] \be \label{ep1.2}
\pi\int\limits_0^{\infty}e^{-\beta U}\;dU= \frac {\pi} {\beta},
\ee where we employ the fact that

\be \label{ep1.3} U=p^2+q^2, \ee and then we use $U$ as a radial
coordinate $U=R^2$, integrate over the angle, and set $dU=2RdR$.
For the mean value of the energy we have
\[<U(p,q)>(\beta)=\frac {1} {Z(\beta)}
\int\limits_{-\infty}^{\infty}H(p,q)e^{-\beta H(p,q)}\;dpdq=\] \be
\label{ep1.4} \frac {\pi} {Z(\beta)} \int\limits_0^{\infty}
Ue^{-\beta U}\;dU=\frac {\pi} {\beta^2 Z(\beta)}, \ee and for the
entropy
\[S(\beta)=\frac {1} {Z(\beta)} \int\limits_{-\infty}^{\infty}
[\ln Z(\beta) + \beta H(p,q)]e^{-\beta H(p,q)}\;dpdq=\] \be
\label{ep1.5} \frac {\pi} {Z(\beta)}  \int\limits_0^{\infty} \{\ln
[Z(\beta)] + \beta U\}e^{-\beta U}\;dU= \frac {\pi} {\beta
Z(\beta)}\{\ln [Z(\beta)] + 1.\} \ee Note  that the integrands
appearing in (\ref{ep1.2}),(\ref{ep1.4}), and (\ref{ep1.5}) are
exact differentials.

\setcounter{equation}{0}

\section{Path Entropy}

\nd Path entropies (phase space curves) have been discussed
recently in Ref. \cite{path,dewar}, for instance. We will be
concerned  here with a {\it related but not identical notion} and
deal with a particle moving in phase space, focusing attention on
its entropy
 evaluated as it moves along some phase space path $\Gamma$. The usefulness of such construct will become evident in the forthcoming Sections.
 Also, as we will
show below, some of the associated paths are adiabatic.

\nd Accordingly, our purpose in this section is to define the
thermodynamic variables of Section 2 {\it on phase-space curves}.
It will be shown that this endeavor is useful. Thus, generalizing
(\ref{ep1.2}),(\ref{ep1.4}), and (\ref{ep1.5}) to curves $\Gamma$,
we define

\begin{itemize}

\item  The partition function as a function of $\beta$ and of a
curve $\Gamma$

\be \label{ep2.1} Z(\beta,\Gamma)=\pi\int\limits_{\Gamma}
e^{-\beta U(p,q)}\;dU(p,q). \ee

\item  The mean energy as

\be \label{ep2.2} <U(p,q)>(\beta,\Gamma)=\frac {\pi}
{Z(\beta,\Gamma)} \int\limits_{\Gamma} U(p,q) e^{-\beta
U(p,q)}\;dU(p,q). \ee

\item   Our path  entropy is defined according to

\be \label{ep2.3} S(\beta,\Gamma)=\frac {\pi}
{Z(\beta,\Gamma)}\int\limits_{\Gamma} \{\ln [Z(\beta,\Gamma)]
+U(p,q)\} e^{-\beta U(p,q)}\;dU(p,q).   \ee

\end{itemize}

\nd   We consider curves, parameterized as a function of the
independent variable $q$, passing through the origin, for which we
have $p(0)=0$ and $q=0$ and as a consequence $U(0,0)=0$. This can
always be the case after an adequate coordinates-change. Moreover,
if we take into account that i) the integrands are exact
differentials and ii) the  integrals are independent of the
curve's shape and only depend on their end-points $q_0$, we have

1) For the partition function
\[Z(\beta,q_0)=\pi\int\limits_0^{q_0}
e^{-\beta U[p(q),q]}\;dU[p(q),q]\] and evaluating the integral \be
\label{ep2.4} Z(\beta,q_0)=\frac {\pi} {\beta}\{1-e^{-\beta
U[p(q_0),q_0]}\}. \ee 2) For the mean value of the energy \be
\label{ep2.5} <U(p,q)>(\beta,q_0)=\frac {\pi}
{Z(\beta,q_0)}\int\limits_0^{q_0} U[p(q),q] e^{-\beta
U[p(q),q]}\;dU[p(q),q], \ee which gives
\[<U(p,q)>(\beta,q_0)=-\frac {\pi} {\beta Z(\beta,q_0)}U[p(q_0),q_0]
e^{-\beta U[p(q_0,q_0]}+\] \be \label{ep2.6} \frac {\pi} {\beta^2
Z(\beta,q_0)}\{1-e^{-\beta U[p(q_0),q_0]}\}. \ee 3) For the
entropy \be \label{ep2.7} S=\frac {\pi}
{Z(\beta,q_0)}\int\limits_0^{q_0} \{\ln Z(\beta,q_0) +U[p(q),q]\}
e^{-\beta U[p(q),q]}\;dU[p(q),q], \ee whose result is
\[S(\beta,q_0)=\frac {\pi} {\beta Z(\beta,q_0)}
\left\{1-e^{-\beta U[p(q_0),q_0]}\right\} \ln [Z(\beta,q_0)]-
\] \be \label{ep2.8}
 \frac {\pi} {Z(\beta,q_0)}U[p(q_0),q_0]
e^{-\beta U[p(q_0),q_0]}+ \frac {\pi} {\beta
Z(\beta,q_0)}\left\{1-e^{-\beta U[p(q_0),q_0]}\right\}. \ee Note
that when $q_0\rightarrow \infty$ (\ref{ep2.4}), (\ref{ep2.6}),
and (\ref{ep2.8}) reduce to (\ref{ep1.2}), (\ref{ep1.4}), and
(\ref{ep1.5}), respectively. Note again that the integrands in
(\ref{ep2.4}), (\ref{ep2.6}), and (\ref{ep2.8}) are {\it exact
differentials}. We insist on the fact that  i) these integrals
become independent of the path $\Gamma$  (i.e., the same for any
$\Gamma$), and ii) if one redefines the coordinate-system  in such
a way that the starting point of $\Gamma$ coincides with the
origin, their values {\bf will depend only on the end-point $q_0$
of the path}. Thus, they are functions of the {\bf microscopic}
state (at least for the HO-Hamiltonian, at this stage).
   We can refer to the entropy and the mean energy evaluated above as
 {\it microscopic thermodynamic potentials} (for the HO).

\section{Equipartition}

\nd In order to ascertain that our thermodynamics along
phase-space curves does make physical sense we look now for an
equipartition theorem. We encounter that

\be <q^2>=\int\limits_{-\infty}^{\infty}\frac {q^2} {Z}
 e^{-\beta(p^2+q^2)}\;dp\;dq=
\frac {\pi} {2Z}\int\limits_{0}^{\infty}Ue^{-\beta U}\;dU,   \ee
i.e., along the curve $\Gamma$

\ben & <q^2>(\beta,\Gamma)=\frac {\pi}
{2Z}\int\limits_{\Gamma}Ue^{-\beta U}\;dU= \frac {\pi}
{2Z}\int\limits_{0}^{q_0}Ue^{-\beta U}\;dU=\cr\cr & =
<q^2>(\beta,q_0)=-\frac {\pi} {2\beta Z(\beta,q_0)}U[p(q_0),q_0]
e^{-\beta U[p(q_0,q_0]}+ \cr \cr & + \frac {\pi} {2\beta^2
Z(\beta,q_0)} \{1-e^{-\beta U[p(q_0),q_0]}\}=\frac
{<U>(\beta,q_0)} {2},  \een that is,

\be <q^2>(\beta,q_0)=<p^2>(\beta,q_0)=\frac {<U>(\beta,q_0)} {2}.
\ee that, for $q_0\rightarrow\infty$, gives

\be <q^2>=<p^2>=\frac {<U>} {2}=\frac {1} {2\beta}, \ee that is,
classical equipartition.

\section{Adiabatic paths}

\nd An adiabatic path is one such that  $S=$ constant along it.
Simplifying (\ref{ep2.8}) we obtain

\[S(\beta,q_0)=\ln \left\{\frac {\pi} {\beta}
\left[1-e^{-\beta U[p(q_0),q_0]}\right]\right\}-\] \be
\label{ep3.1} \frac {\beta U[p(q_0),q_0]e^{-\beta U[p(q_0),q_0]}}
{1-e^{-\beta U[p(q_0),q_0]}}+1. \ee The condition $S=$ constant
translates into

\be \label{ep3.2} \beta=C_1\;\;\; U[p(q_0),q_0]=C_2, \,\,\,{\rm
independently  \,\,\,of  \,\,\,q_0}. \ee $C_1=\beta$ is constant
by the very reservoir's notion. For the curve  $p=f(q)$ this
entails, for our Hamiltonian, that

\be  p^2+q^2 = (p + \delta p)^2+(q+ \delta q)^2,  \ee i.e.,

\be p \delta p =- q\delta q.     \ee For the curve $p=f(q)$ one
has $p\delta p= pf'(q)\delta q$ and

\be \label{circle} f(q)f'(q) = - q,  \ee is the equation that
yields an adiabatic path $f(q)$ (indeed, an infinite family of
paths since an integration constant $C$ will emerge in solving the
pertinent equation). The solution of (\ref{circle}) is obtained
after transforming it into

\ben \label{circle1} & \frac{d f^2}{dq} = - 2q, \cr & f(q)^2= -q^2
+C\,\,\, \rightarrow \,\,\, p^2 +q^2 =C,\een which is intuitively
obvious. We may dare to conjecture that for any Hamiltonian of the
form $H=g_1(q)+ g_2(q)$ the end points of the adiabatic paths
might be of the form $g_1(q)+ g_2(q)= constant$.

 \nd A
slightly different question is that of finding two straight-line
paths (passing trough the origin) with the same entropy. They are
found as follows:

 \be \label{ep3.3} p(q)=aq, \ee so that we should have, for two
different lines

\be \label{ep3.4} U=(a^2+1)q_0^2=(a^{'2}+1)q^{'2}_0. \ee If we
take
\[a^{'}<a\]
and \be \label{ep3.5} q_0^{'}=\sqrt{\frac {a^2+1} {a^{'2}+1}}q_0,
\ee then

\be \label{ep3.6} \Delta
S=S(\beta,a^{'},q_0^{'})-S(\beta,a,q_0)=0. \ee If the evolution of
the system starts from the line $p=aq$, ends in the line $p=a^{'}
q^{'}$, and crosses all the space between the two lines then,
whenever (\ref{ep3.2}) is satisfied, the evolution is adiabatic.
\setcounter{equation} {0}

\section{Entropic Force}


\nd We arrive here at our main theme. According to (\ref{ep3.1}),
the entropic force is given by Eq. (3.3) of \cite{verlinde} that
reads  $F_e dx= TdS $. In our case this translates as

\be \label{ep4.1bis } F_e dq=\frac {1} {\beta}\frac {\partial S}
{\partial q} dq, \ee and

\be \label{ep4.1}  F_e= \beta U \frac {\partial U[p(q),q]} {\partial q}
e^{-\beta U} \frac {2-e^{-\beta U}} {(1-e^{\beta U})^2} \ee
where the trajectory's end-point is free to move in phase-space.
For $\beta U<<1$  (the quantum limit), Eq. (\ref{ep4.1})
simplifies to


\be \label{ep4.2} F_e=\frac {\partial U[p(q),q]} {\partial q}
\left\{\frac {1} {\beta U[p(q),q]}-\beta U[p(q),q]\right\}
\ee or


\be \label{ep4.3} F_e= 2q \left\{\frac {1} {\beta U[p(q),q]}-\beta U[p(q),q]\right\}\sim  2q
\frac {1} {\beta U[p(q),q]}. \ee Thus, there is a strong
repulsion. Actually, a hard core at q=0. We are dealing with a
particle attached via spring to the origin, that cannot be reached
due to the entropic force.

\section{Entropic Force on arbitrary phase-space curves}

\nd More generally, for $U=p^2+q^2$ and \color{red} any curve
\normalcolor in phase space, one has

\be \label{general}  F_e=2q \beta (p^2+q^2) e^{-\beta (p^2+q^2)}
\frac {2-e^{-\beta (p^2+q^2)}} {[1-e^{\beta (p^2+q^2)}]^2} \ee \nd
We present 3-dimensional plots and $F_e-$level curves for three
temperature regimes, namely,
\begin{itemize}

\item Low temperatures, $\beta=5$ (Figs. 1-2),

\item Intermediate temperatures, $\beta=1$ (Figs. 3-4),

\item High temperatures, $\beta=0.2$  (Figs. 5-6).

\end{itemize}
\nd We see that there is an infinitely repulsive barrier (hard
core) near (but not {\bf at}) the origin. In the immediate
vicinity of the origin the force vanishes. It also tends to zero
at long distances from the hard-core. The conjunction between
these facts yields both confinement and asymptotic freedom via a
simple classical mechanism.




\section{The total well that our particle feels}

\nd Of course, our particle not only feels the $F_e-$influence but
also that of the negative gradient of the HO potential . Thus, it
is affected by a total force $F_{Tot}= F_e+F_{HO}$. The pertinent
expression

\be F_T=q[1+3\beta(p^2+q^2)
-e^{-\beta(p^2+q^2)}-2\beta(p^2+q^2)e^{-\beta(p^2+q^2)}] \frac
{e^{-\beta (p^2+q^2)}} {[1-e^{-\beta (p^2+q^2)}]^2},\ee where \be
F_{HO}=q[1-\beta(p^2+q^2)-e^{-\beta(p^2+q^2)}] \frac
{e^{-\beta(p^2+q^2)}} {[1-e^{-\beta(p^2+q^2)}]^2}. \ee We plot
this total force for, respectively, $\beta = 0.2, 1.0,$ and $5.0$
in Figs. 7, 8,  and 9. It is seen that the essential features
described in the preceding Section do not suffer any appreciable
qualitative change.


\section{Clausius relation and specific heat}

\nd Let us now consider, for an infinitesimal work $dW$ generated
by a change $dq_0$

\[d<U>=TdS - dW,\]
where $dW$ is the work done ON the system if $dq_0 < 0$
\cite{deslog}. In one dimension, the pressure reduces, of course,
to a force. One obtains

\be  dW=\frac {e^{-\beta U[p(q),q]}} {1-e^{-\beta U[p(q),q]}},\ee
and, according to
\[ dW= F dq,\]
for the linear force (pressure in one dimension) $F_{linear}$ we
see that it is $\Gamma-$dependent and given by

\be F_{linear}(\Gamma) =\frac {e^{-\beta (p^2+q^2)}\left(2p\frac
{dp} {dq}+ 2q\right)} {1-e^{-\beta (p^2+q^2)}}, \label{linearF}
\ee that, we insist,  depends on the curve $\Gamma$ [remember that
$p$ and $q$ possess common dimensionality (see Eq. (1))]. Figs.
10, 11, and 12 depict $F_{linear}$ for, respectively, $\beta=$
=.2, 1, and 5, with $\Gamma$ being given by $p=-q^2+q$. The force
vanishes almost everywhere. There is a clear transition near the
hard core and, significantly enough,  it becomes negative on on
side of it. Now, negative pressures (linear force in our case) are
a distinctive property of dark energy, a hypothetical form of
energy that permeates all of space and tends to accelerate the
expansion of the universe \cite{dark}. Indeed, it constitutes the
most accepted hypothesis to explain observations dating from the
90's that indicate that the universe is expanding at an
accelerating rate. Note here that, independently from its actual
nature, dark energy would need to have a strong negative pressure
(acting repulsively) in order to explain the observed acceleration
in the expansion rate of the universe. According to General
Relativity, the pressure within a substance contributes to its
gravitational attraction for other things just as its mass density
does. This happens because the physical quantity that causes
matter to generate gravitational effects is the stress-energy
tensor, which contains both the energy (or matter) density of a
substance and its pressure and viscosity. \vskip 3mm

\nd  Finally, the {\it specific heat}  is easily seen to be

\be C=k_{Boltzmann} \left\{1-\frac
{\beta^2(p^2+q^2)e^{-\beta(p^2+q^2)}}
{\left[1-e^{-\beta(p^2+q^2)}\right]^2}\right\},\ee independently
of the curve $\Gamma$. Figs. 13, 14, and 15
 depict $C$ for, respectively, $\beta=$ =.2, 1, and 5.  The hard core generates a phase transition. The specific heat
 changes sign and becomes negative near it, and  drops
 rapidly  near the origin. Negative
 specific heats are perhaps the most distinctive  thermodynamic feature of
 self-gravitating systems \cite{binney}. Here, our entropic discourse
 establishes    thereby
 contact with Verlinde's work [3].

\section{Discussion}

\nd We were dealing with a particle attached to the origin by a
spring and consider entropic-force effects. Although we focus
attention upon arbitrary phase space curves $\Gamma$, most of our
effects were independent of the specific path $\Gamma$.
 Our statistical mechanics-along-curves concept is seen to make sense because
 the equipartition theorem is valid for it.

 \nd We
considered the entropic construct of Eq. (\ref{ep2.2}) and we saw
that the equipartition theorem holds. From Figs. 1-6 we gather the
entropic force diverges at short distances from the origin
(hard-core effect), but vanishes both just there
 and at infinity, so that, with some abuse of language one may speak of ``asymptotic freedom".
\nd The entropic force is repulsive.  As stated above, at long
distances  from the origin the entropic force tends to vanish. The
negative specific heat we encounter near the hard core links our
work to that of Verlinde's [3].

\vskip 3mm

\nd Entropic confinement is the most remarkable effect that our
classical entropic force-model exhibits. Independently  of whether
our model is realistic or not, it does provide a classical
confinement mechanism. The present considerations should encourage
non-classical explorations regarding the entropic force. \vskip
3mm

\nd Finally, when we couple the entropic force effects with those
of the HO-potential we are not  able to discern significant new
features.

\newpage

\begin{figure}[h!]
\begin{center}
\includegraphics[width=8.6cm]
{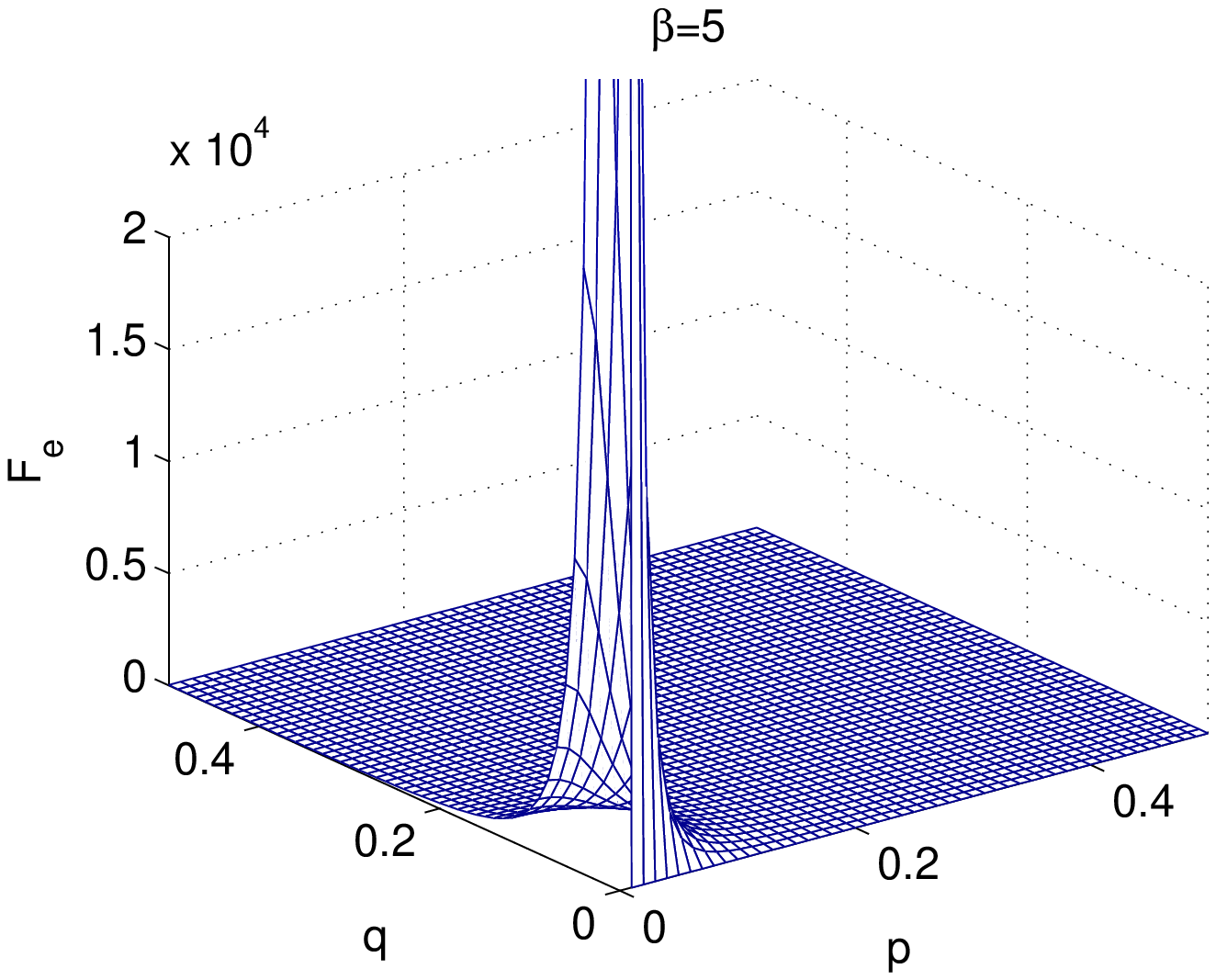} \caption{Arbitrary curves on phase-space.
  Entropic force vs. q, p for $\beta=5$ (low temperature). Note the hard-core barrier and
the vanishing of the in a neighborhood of  the origin.}
\label{figura1}
\end{center}
\end{figure}

\begin{figure}[h!]
\begin{center}
\includegraphics[width=8.6cm]
{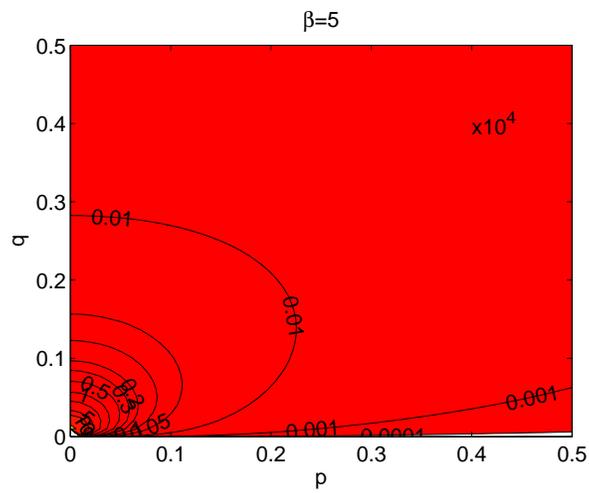} \caption{Arbitrary curves on phase-space.
Level $F_e-$curves in the q-p plane  (low temperature,
$\beta=5$).} \label{figura2}
\end{center}
\end{figure}

\begin{figure}[h!]
\begin{center}
\includegraphics[width=8.6cm] {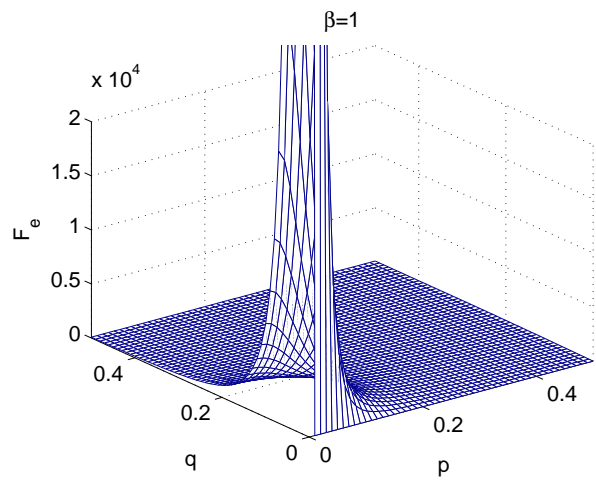}
\caption{Arbitrary curves on phase-space.
  Entropic force vs. q, p for $\beta=1$ (intermediate temperature). Note the hard-core barrier and
the vanishing of the in a neighborhood of  the origin.}
\label{figura3}
\end{center}
\end{figure}

\begin{figure}[h!]
\begin{center}
\includegraphics[width=8.6cm]
{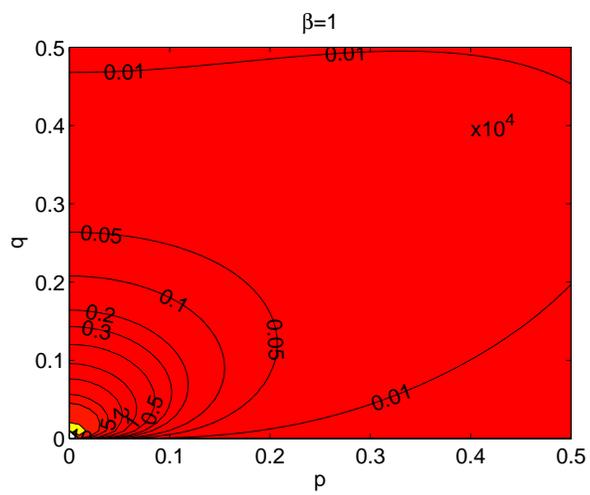} \caption{Arbitrary curves on phase-space.
Level $F_e-$curves in the q-p plane(intermediate temperature,
$\beta=1$ ).} \label{figura4}
\end{center}
\end{figure}

\begin{figure}[h!]
\begin{center}
\includegraphics[width=8.6cm]
{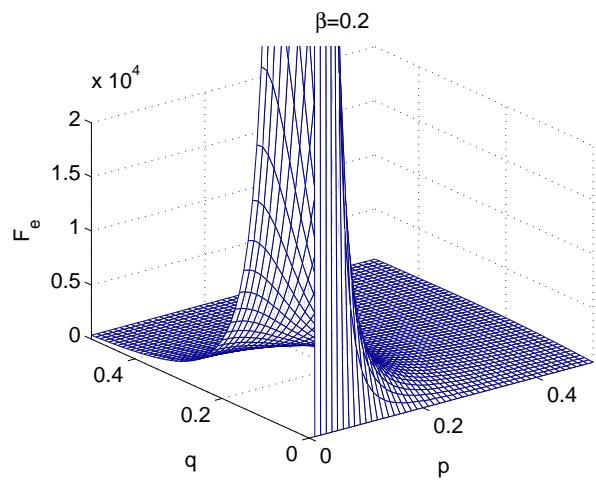} \caption{Arbitrary curves on phase-space.
Entropic force vs. q, p for $\beta=0.2$ (high temperature). Note
the hard-core barrier, the vanishing of the force at the origin
and the attraction/repulsion zones.} \label{figura5}
\end{center}
\end{figure}

\begin{figure}[h!]
\begin{center}
 \includegraphics[width=8.6cm]{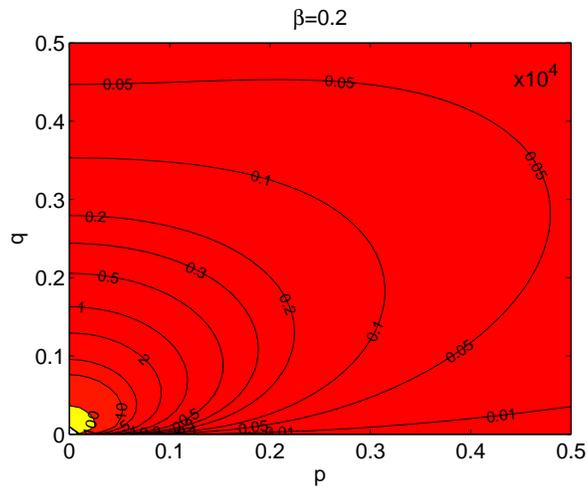}\caption{Arbitrary curves on phase-space.
  Level $F_e-$curves in the q-p plane for $\beta=0.2$ (high temperature). }
\label{figura6}
\end{center}
\end{figure}

\begin{figure}[h!]
\begin{center}
\includegraphics[width=8.6cm]
{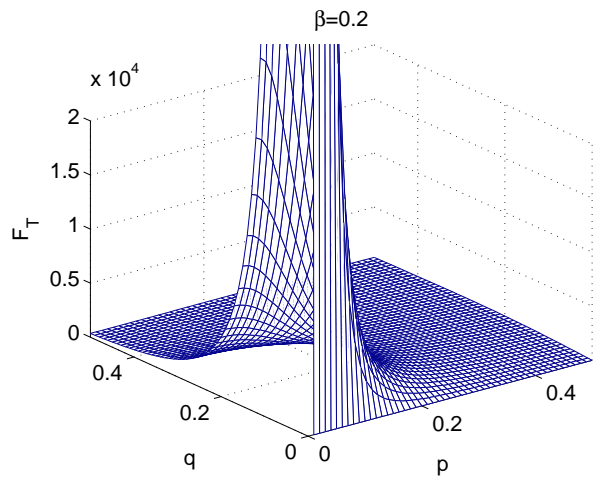} \caption{Total force $F_T$ for $\beta=0.2$.}
\label{figura7}
\end{center}
\end{figure}

\begin{figure}[h!]
\begin{center}
\includegraphics[width=8.6cm]
{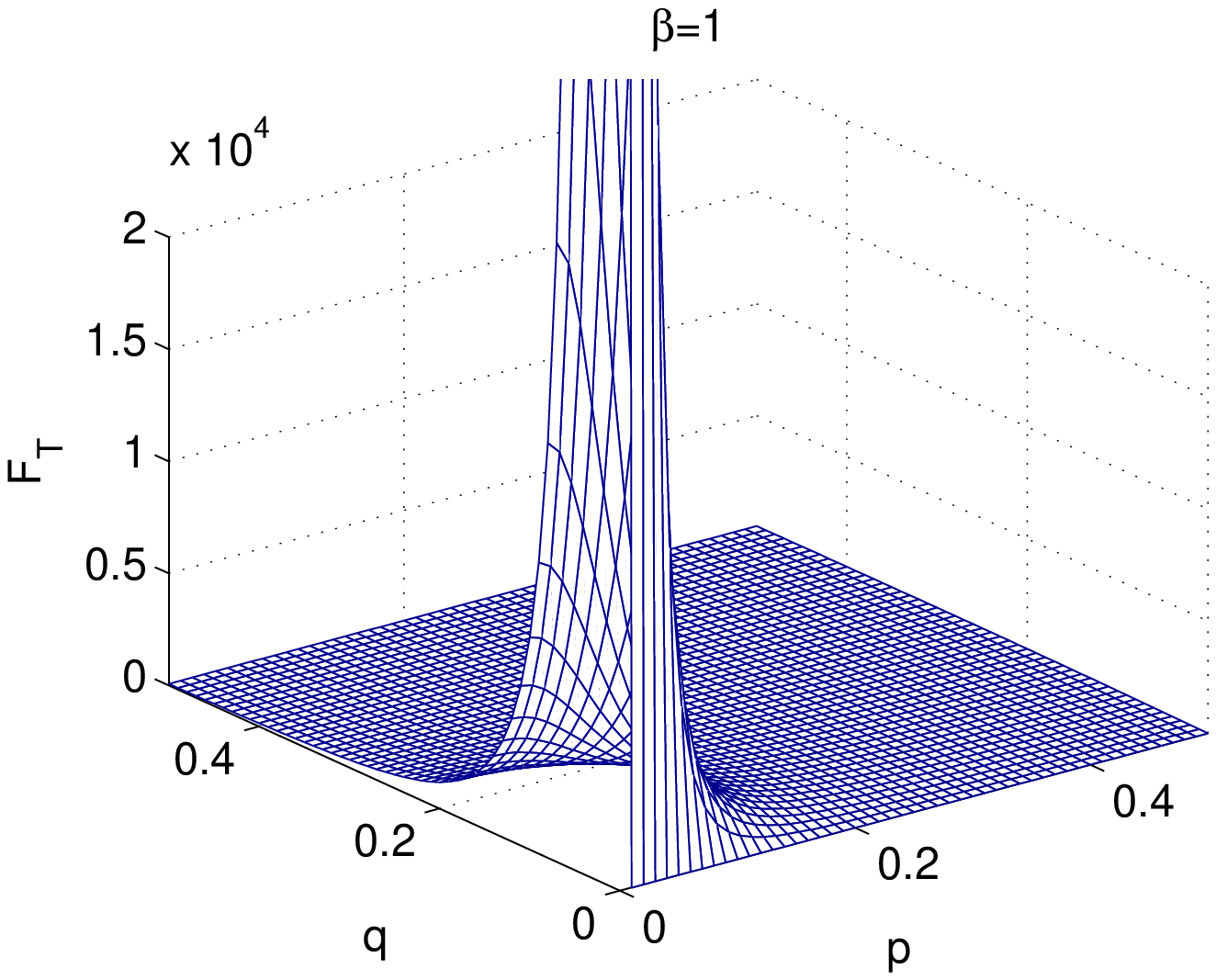} \caption{Total force $F_T$ for $\beta=1.0$.}
\label{figura8}
\end{center}
\end{figure}

\begin{figure}[h!]
\begin{center}
\includegraphics[width=8.6cm]
{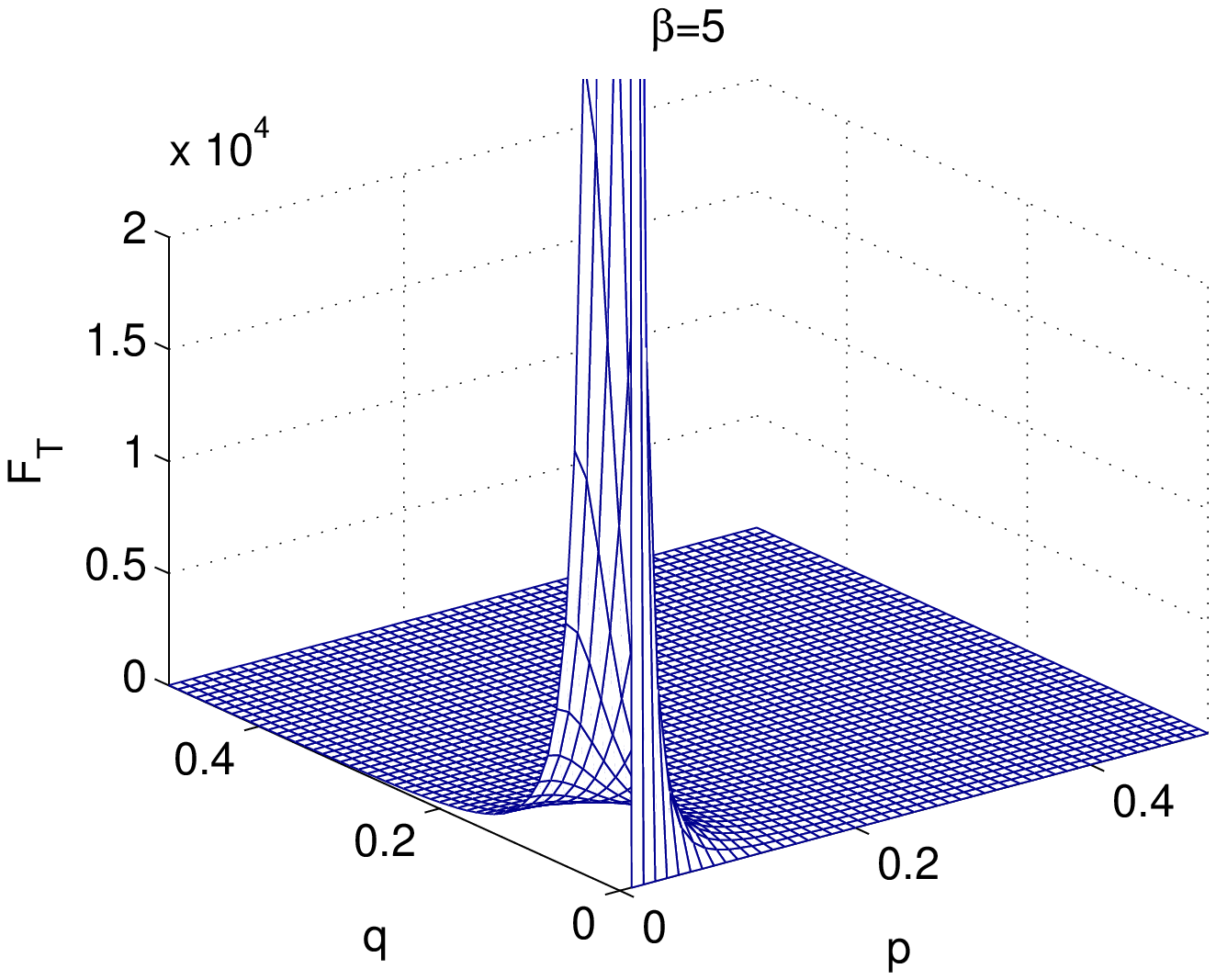} \caption{Total force $F_T$ for $\beta=5.0$.}
\label{figura9}
\end{center}
\end{figure}

\begin{figure}[h!]
\begin{center}
\includegraphics[width=8.6cm]
{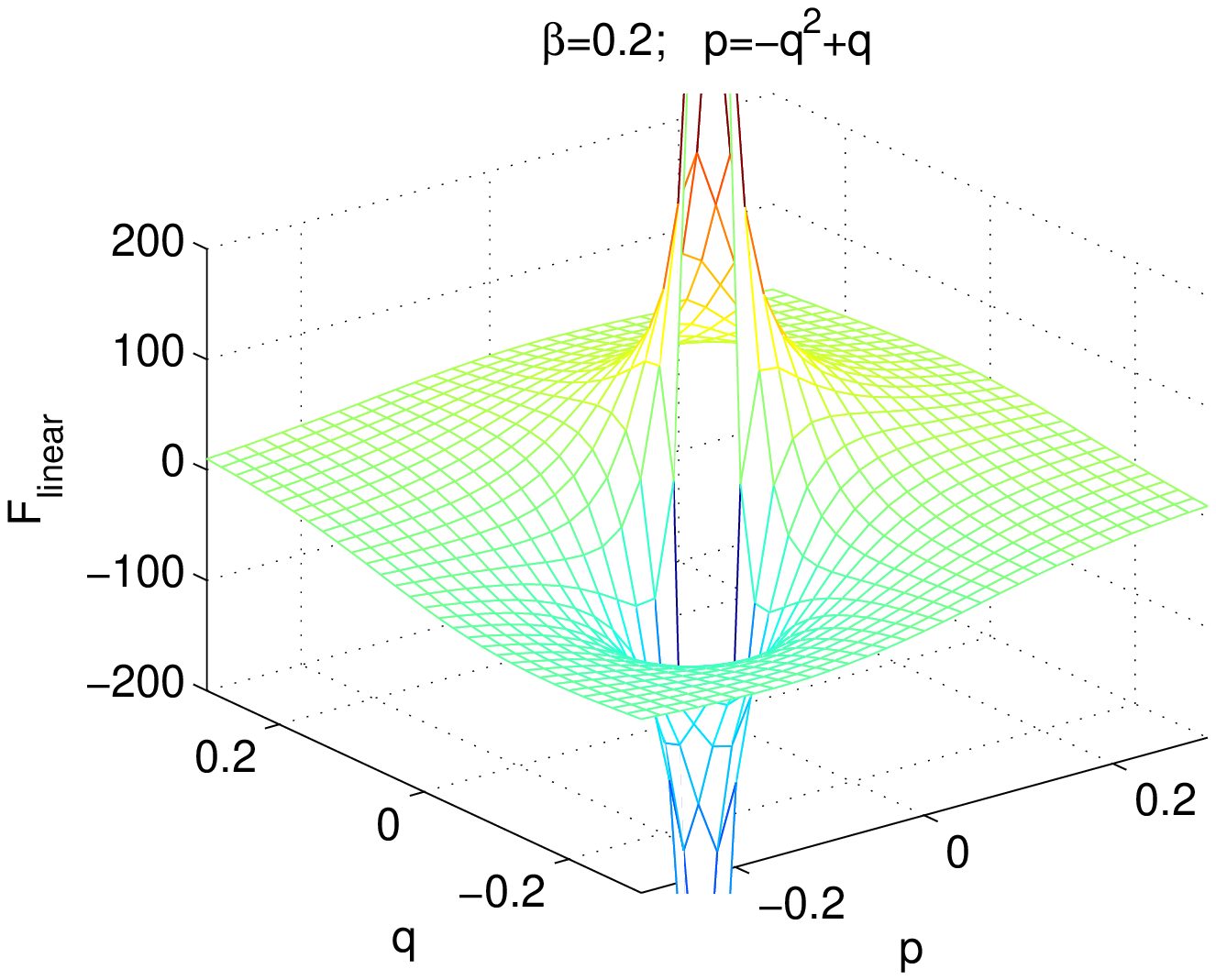} \caption{An example of the linear  force
$F_L$'s behavior for $\beta=0.2$.} \label{figura10}
\end{center}
\end{figure}

\begin{figure}[h!]
\begin{center}
\includegraphics[width=8.6cm]
{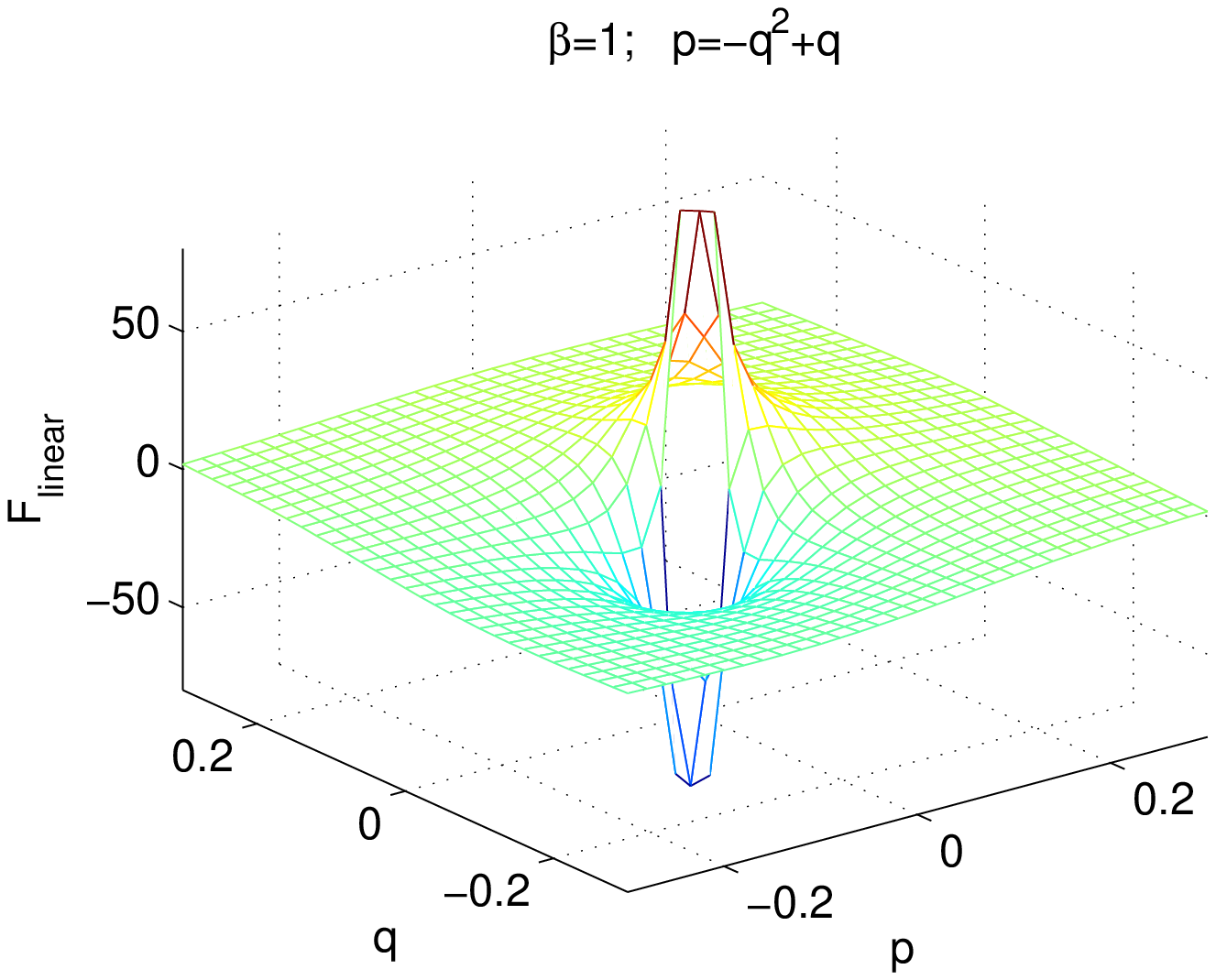} \caption{An example of the linear
force $F_L$'s behavior for   $\beta=1.0$.} \label{figura11}
\end{center}
\end{figure}

\begin{figure}[h!]
\begin{center}
\includegraphics[width=8.6cm]
{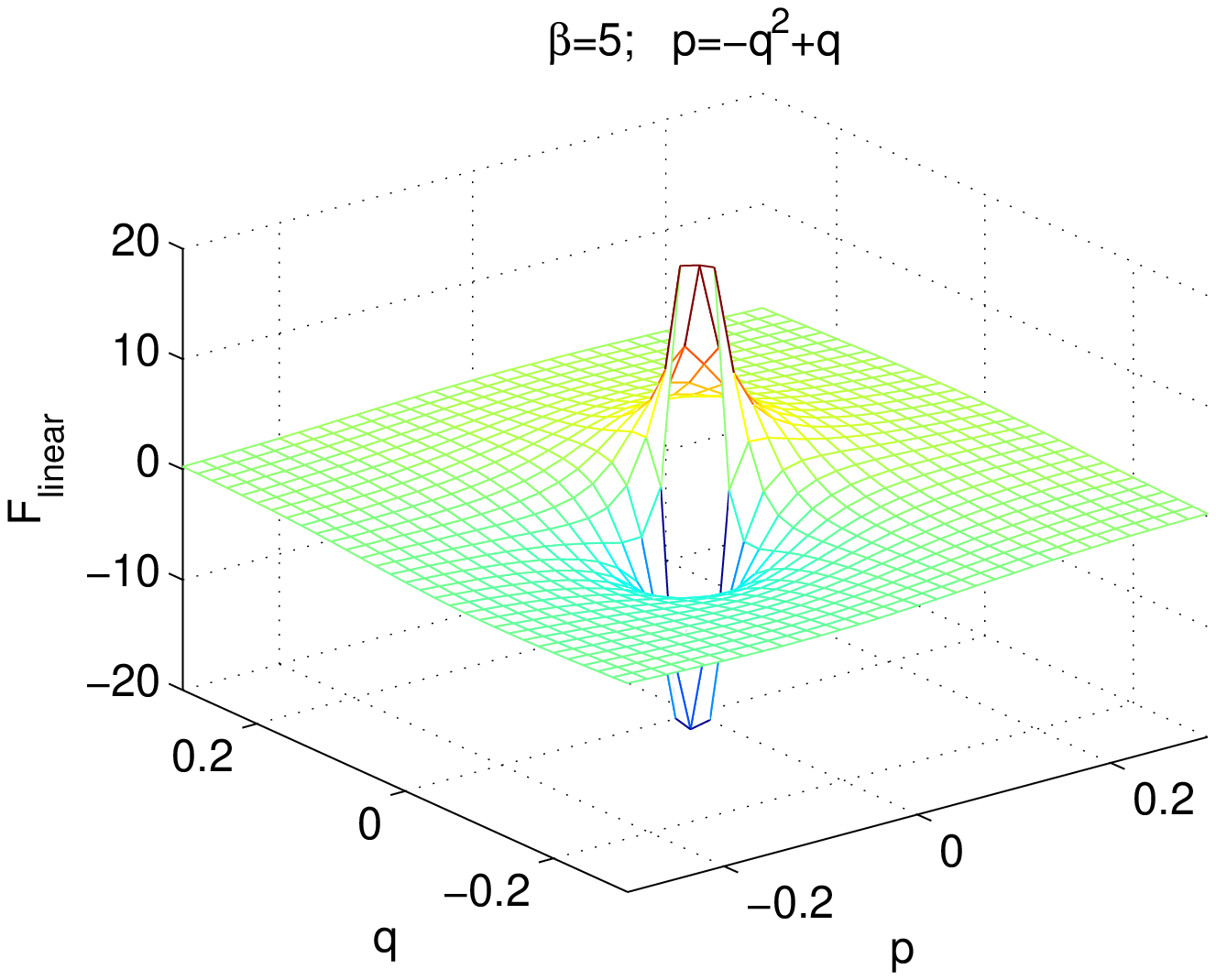} \caption{An example of the linear  force
$F_L$'s behavior for $\beta=5.0$.} \label{figura12}
\end{center}
\end{figure}

\begin{figure}[h!]
\begin{center}
 \includegraphics[width=8.6cm]{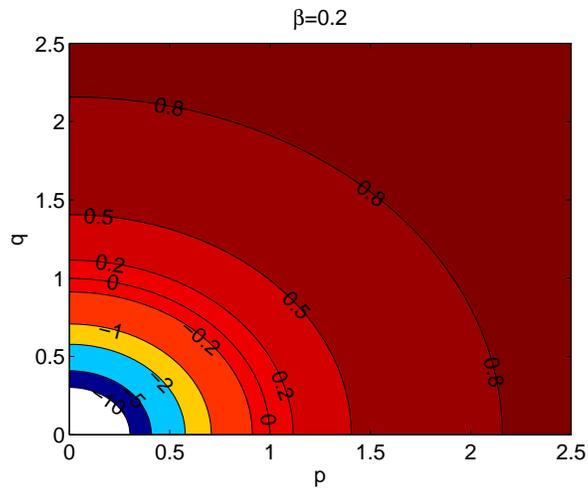}\caption{Specific heat.
  Level $C-$curves  in the q-p plane for $\beta=0.2$ (high temperature). }
\label{figura13}
\end{center}
\end{figure}

\begin{figure}[h!]
\begin{center}
 \includegraphics[width=8.6cm]{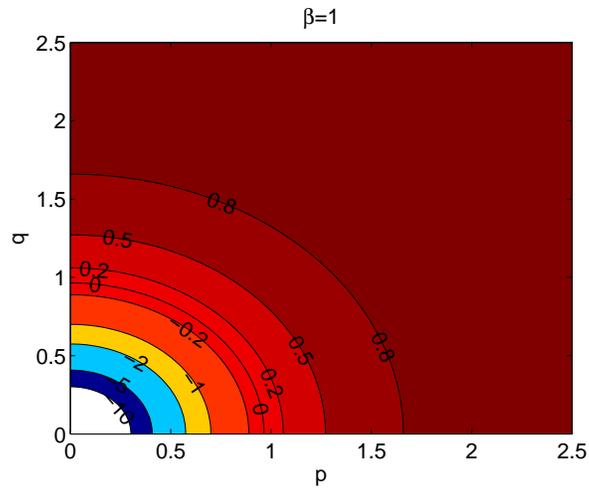}\caption{
  Level $C-$curves in the q-p plane for $\beta=1.0$ (intermediate temperature). }
\label{figura14}
\end{center}
\end{figure}

\begin{figure}[h!]
\begin{center}
 \includegraphics[width=8.6cm]{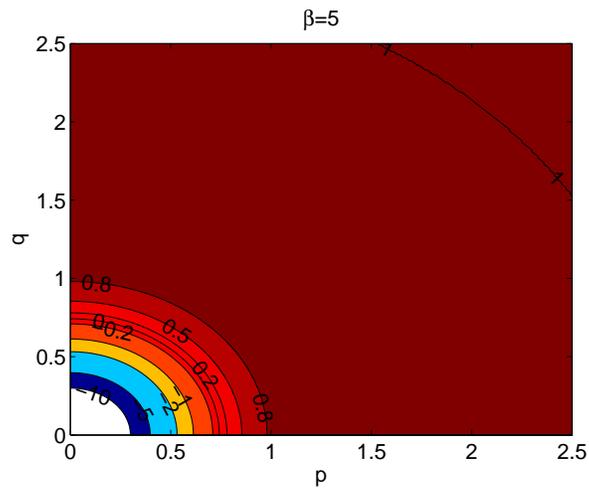}\caption{
  Level $C-$curves in the q-p plane for $\beta=5.0$ (low temperature). }
\label{figura15}
\end{center}
\end{figure}

  \end{document}